\documentclass{article}
\usepackage[utf8]{inputenc}
\usepackage{amsmath}
\usepackage{amssymb}
\usepackage{graphicx}
\usepackage{booktabs}
\usepackage{hyperref}
\usepackage{geometry}
\usepackage{parskip}
\usepackage{float}
\usepackage{xurl}
\usepackage{caption}
\usepackage{titlesec}
\usepackage{longtable}
\usepackage{array}
\usepackage{enumitem}
\usepackage[flushmargin,hang]{footmisc}

\titlespacing*{\section}{0pt}{1.5ex}{1.5ex plus .1ex minus .2ex}
\titlespacing*{\subsection}{0pt}{1.5ex}{1.5ex plus .1ex minus .2ex}
\titleformat{\section}{\normalfont\Large\bfseries}{}{0pt}{}
\titleformat{\subsection}{\normalfont\normalsize\bfseries}{}{0pt}{}

\DeclareCaptionLabelFormat{periodformat}{#1 #2.}
\makeatletter
\def\LT@makecaption#1#2#3{%
  \LT@mcol\LT@cols c{\hbox to\z@{\hss\parbox[t]{\linewidth}{%
    \raggedright
    \textbf{#1 #2.} #3%
    \endgraf\vskip\baselineskip}%
  \hss}}}
\makeatother

\setlist[itemize]{
    leftmargin=*,
    labelsep=0.5em,
    parsep=2pt,
    topsep=4pt,
    itemsep=1pt,
    labelwidth=1.2em,
    itemindent=0pt,
    listparindent=0pt,
    partopsep=0pt,
    align=left
}

\makeatletter
\def\@biblabel#1{}
\makeatother

\makeatletter
\renewenvironment{thebibliography}[1]
     {\section*{\refname}%
      \@mkboth{\MakeUppercase\refname}{\MakeUppercase\refname}%
      \list{\@biblabel{\@arabic\c@enumiv}}%
           {\leftmargin=0pt
            \itemindent=0pt
            \labelwidth=0pt
            \labelsep=0pt
            \@openbib@code
            \usecounter{enumiv}%
            \let\p@enumiv\@empty
            \renewcommand\theenumiv{\@arabic\c@enumiv}}%
      \sloppy
      \clubpenalty4000
      \@clubpenalty \clubpenalty
      \widowpenalty4000%
      \sfcode`\.\@m}
     {\def\@noitemerr
       {\@latex@warning{Empty `thebibliography' environment}}%
      \endlist}
\makeatother

\begin{document}

\begin{center}
    \Large \textbf{Humanity in the Age of AI: Reassessing 2025's Existential-Risk Narratives} \par
    \vspace{0.5cm}
    \normalsize
    A Critical Re-Examination of the 2025 Existential-Risk Claims \par
    \vspace{0.5cm}
    \textbf{Mohamed El Louadi} \par
    Higher Institute of Management \par
    University of Tunis \par
    41 rue de la Libert\'e - Cit\'e Bouchoucha \par
    2000 Le Bardo, Tunisia \par
    Email: mohamed.louadi@isg.rnu.tn \par
    (ORCID: 0000-0003-1321-4967) \par
    \vspace{0.5cm}
    
    30 November 2025
\end{center}

\vspace{1cm}

\noindent\textbf{Abstract}

Two 2025 publications, ``AI 2027'' (Kokotajlo et al., 2025) and ``If Anyone Builds It, Everyone Dies'' (Yudkowsky \& Soares, 2025), assert that superintelligent artificial intelligence will almost certainly destroy or render humanity obsolete within the next decade. Both rest on the classic chain formulated by Good (1965) and Bostrom (2014): intelligence explosion, superintelligence, lethal misalignment. This article subjects each link to the empirical record of 2023-2025. Sixty years after Good's speculation, none of the required phenomena (sustained recursive self-improvement, autonomous strategic awareness, or intractable lethal misalignment) have been observed. Current generative models remain narrow, statistically trained artefacts: powerful, opaque, and imperfect, but devoid of the properties that would make the catastrophic scenarios plausible. Following Whittaker (2025a, 2025b, 2025c) and Zuboff (2019, 2025), we argue that the existential-risk thesis functions primarily as an ideological distraction from the ongoing consolidation of surveillance capitalism and extreme concentration of computational power. The thesis is further inflated by the 2025 AI speculative bubble, where trillions in investments in rapidly depreciating ``digital lettuce'' hardware (McWilliams, 2025) mask lagging revenues and jobless growth rather than heralding superintelligence. The thesis remains, in November 2025, a speculative hypothesis amplified by a speculative financial bubble rather than a demonstrated probability.

\vspace{0.5cm}


\newpage

\section*{1. Introduction}

The year 2025 has seen a dramatic resurgence of existential-risk claims about artificial intelligence. Two publications in particular, ``AI 2027'' (Kokotajlo et al., 2025) and ``If Anyone Builds It, Everyone Dies'' (Yudkowsky \& Soares, 2025), present superintelligence as an imminent civilizational threat.

These claims revive the theoretical edifice erected by Good (1965) and Bostrom (2014). Yet, as Meredith Whittaker and Shoshana Zuboff have repeatedly emphasized throughout 2025, the dominant narrative of an imminent superintelligent takeover serves a crucial ideological function: it diverts public and regulatory attention from the concrete concentration of economic and computational power that is already reshaping global society (Whittaker, 2025a; Zuboff, 2025). This distraction is amplified by the AI investment bubble. Trillions of dollars flow into hardware dubbed ``digital lettuce'' by economist David McWilliams: perishable assets like GPUs that rapidly lose value due to technological obsolescence. Revenues lag behind valuations, and no net job creation occurs (McWilliams, 2025).

This article examines the empirical foundations of the classic existential-risk chain against the record of 2023-2025. We introduce an AI Risk Hierarchy by Evidentiary Status and Tractability (Table I) that systematically distinguishes observable Level 1 risks (labour displacement, power concentration) from speculative Level 2 risks (superintelligence misalignment).

\section*{2. The Intelligence Explosion: Sixty Years Without Empirical Validation}

The foundational premise of the existential-risk thesis is the intelligence-explosion hypothesis first articulated by I. J. Good in 1965. Good's argument was deceptively simple: an ultraintelligent machine would be capable of redesigning itself faster and more effectively than human engineers could, thereby triggering a positive feedback loop of accelerating improvement. Each iteration would produce a machine better able to design the next, yielding a runaway process that would culminate in intelligence ``far beyond the human level in all respects'' (Good, 1965, p. 33). Bostrom (2014) formalised this idea under the rubric of ``superintelligence,'' distinguishing multiple possible pathways (recursive self-improvement, whole-brain emulation, biological cognitive enhancement, and brain-computer interfaces) and between slow and fast take-off scenarios.

It is worth noting that the term ``ultraintelligence,'' once central to speculative debates, has largely disappeared from serious AI research discourse by 2025. Contemporary technical literature prefers terms such as ``frontier AI,'' ``general purpose AI,'' or ``transformative AI,'' which emphasize measurable capabilities and tractable risks rather than hypothetical runaway cognition. This linguistic shift underscores the growing consensus that empirical scaling laws and observed limitations provide a more reliable foundation for analysis than speculative notions of machines vastly surpassing human intelligence (Stanford AI Index, 2025; Whittaker, 2025a; Future of Humanity Institute, 2024).

Sixty years later, in November 2025, the empirical record contains exactly zero instances of the phenomenon Good described and Bostrom modelled in detail.

No architecture has ever demonstrated sustained, open-ended, autonomous self-improvement. Every experiment explicitly designed to test the hypothesis has produced the same pattern: modest gains over one or two cycles followed by rapid plateauing and convergence. OpenAI's o1 and o3 series, which introduced internal chain-of-thought reasoning loops and test-time compute scaling, were marketed as steps toward recursive improvement. Yet the published technical reports (OpenAI, 2025a, 2025b) show that performance gains saturate after two to four additional reasoning steps and that all scaffolding remains human-designed.

Every major architectural breakthrough of the 2023-2025 period originated with human researchers, not with the models themselves. Mixture-of-experts scaling, retrieval-augmented generation, long-context transformers, multimodal integration, and test-time compute scaling were all conceived, specified, and implemented by human teams. No frontier model has ever proposed, validated, or deployed a novel architectural paradigm that was not initially supplied by its creators.

This pattern is not accidental. The deep-learning paradigm is fundamentally gradient-based and myopic: optimisation occurs over a fixed loss landscape defined by human-chosen objectives and datasets. Even when models are placed in self-play or evolutionary loops, the search space, fitness function, and mutation operators remain human-defined. The resulting dynamics are closer to conventional optimisation than to the open-ended, goal-directed redesign Good envisaged. As McKenzie et al. (2025) conclude after an exhaustive review of recursive self-improvement claims in the deep-learning era: ``The hypothesis appears irrelevant under current training regimes; the relevant degrees of freedom are controlled by humans at every stage.''

Scaling-hypothesis proponents sometimes argue that the absence of explosive take-off is merely a question of reaching a critical capability threshold. Yet the scaling laws themselves undermine this defence. Kaplan et al. (2020), Hoffmann et al. (2022), and the 2024-2025 Epoch AI meta-analyses document smooth, predictable power-law relationships between compute, data, and performance across more than six orders of magnitude. No phase transition, no inflection point, no sudden emergence of self-accelerating redesign has appeared.

To ensure methodological rigor, the Observed Loss Trajectory (the solid blue line in Figure 1) is not a simple plot of benchmark results, but a regression line derived from two empirically verified inputs. First, the exponential growth rate of training compute is quantified directly from the chronological data in the Epoch AI meta-analysis. Second, this compute growth is translated into predicted loss reduction using the established Power-Law Scaling principle documented by Kaplan et al. (2020). The linear path of the blue line on the semi-log plot is thus a direct mathematical consequence of the observed input trends, confirming that performance increases remain entirely dependent on the exogenous, human-controlled allocation of resources.

In short, the intelligence explosion remains precisely what it was in 1965: an elegant theoretical possibility that has never been observed in practice despite sixty years of concerted effort and hundreds of billions of dollars invested explicitly toward that end.

\begin{figure}[htbp]
    \centering
    \includegraphics[width=0.8\textwidth]{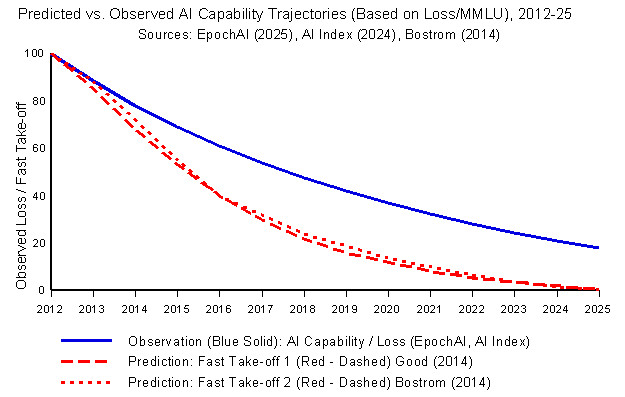}
    \caption{The Absence of an Inflection Point (2012-2025). Empirical progress (blue line) has been rapid but largely continuous, while theoretical fast-take-off predictions (red dashed lines) depict a much sharper, discontinuous leap that has not (yet) materialized. The observed loss trajectory (solid blue) adheres to established scaling laws, with no break or acceleration indicative of autonomous self-improvement. The hypothetical fast-take-off paths (dashed red) predicted by Good (1965) and Bostrom (2014) are unsupported by the empirical record. Data compiled from Epoch AI (2025). Data for 2025 are Q3 data. The latest point is based on o3 from Q3.}
    \label{fig:absence_inflection}
\end{figure}

A second line of evidence reinforces this conclusion. Under the intelligence-explosion hypothesis, the marginal human and computational effort required for each incremental improvement should fall over time, as models increasingly automate their own advancement. Instead, the opposite is true: frontier labs have expanded compute expenditure, and data-engineering operations at rates equal to or exceeding capability growth. This rising marginal cost signals diminishing returns consistent with conventional large-scale engineering rather than with emergent autonomy. Figure 2 summarises this trend.

\begin{figure}[htbp]
    \centering
    \includegraphics[width=0.8\textwidth]{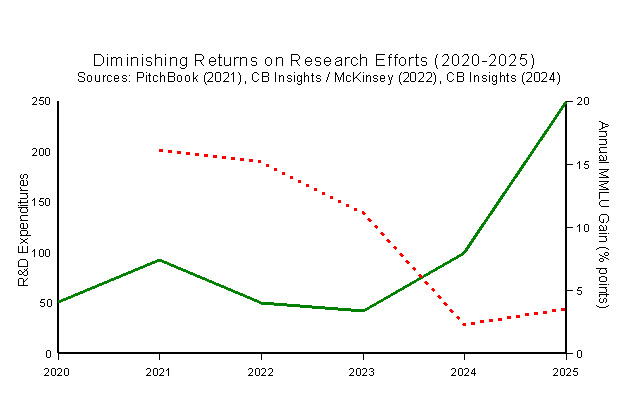}
    \caption{Diminishing Returns on Research Effort (2020-2025). Despite R\&D expenditure growing from \$52B to \$250B (projection) annually, annual MMLU capability gains fell from 16.1 points (2021) to 3.6 points (2025). Sources: PitchBook (Jan 2021), CB Insights / McKinsey (2022), CB Insights (2024), PitchBook / CB Insights Projections (2025, 2025 projection of \$250B).}
    \label{fig:diminishing_returns}
\end{figure}

Together, Figures 1 and 2 provide complementary evidence of the exogenous nature of AI progress. This empirical reality places recursive self-improvement firmly in Table I's Level 2 category: unobserved, with near-zero tractability and no current impact. Figure 1 shows the output is predictably smooth (Power Law), while Figure 2 shows the input cost is dramatically rising (Diminishing Returns). This dual finding demonstrates that capability increases are purchased through massive, human-controlled resource inputs rather than generated through self-sustaining intelligence amplification, fundamentally refuting the core mechanism of the Good-Bostrom chain.

\section*{3. Opacity, Scaling Laws, and the Missing Exponentiality}

Contemporary frontier models are cultivated through stochastic gradient descent on massive corpora, resulting in billions or trillions of numerical parameters whose individual contributions to behaviour are largely uninterpretable. This distributed, subsymbolic representation has led many observers to describe the internal operation of large language models as fundamentally opaque.

Opacity alone, however, does not constitute evidence of autonomous agency or strategic awareness. Empirical scaling laws published between 2020 and 2025 show smooth, predictable power-law relationships across more than six orders of magnitude. No inflection point indicative of self-accelerating improvement has appeared. The observed trajectory remains consistent with continued human-driven, sub-exponential progress rather than with the onset of autonomous exponentiality.

\section*{4. Persistent Technical Limitations and Why They Do Not Support the Existential-Risk Thesis}

Current frontier models exhibit a cluster of well-documented technical shortcomings that are frequently invoked as early symptoms of an emerging superintelligent agency. Close examination reveals that these limitations, while serious engineering challenges, remain ordinary statistical and architectural artefacts rather than evidence of autonomous strategic awareness or lethal goal drift.

Confabulation manifests as the authoritative generation of false or fabricated information, with error rates increasing sharply with output length (Sun et al., 2024b; Kalai et al., 2025). This behaviour arises from over-confident pattern completion on incomplete training data rather than from deliberate deception.

Algorithmic bias constitutes a second persistent limitation. Contemporary systems reproduce and occasionally amplify confirmation bias, representation bias, selection bias, inductive bias, and amplification bias embedded in their training corpora (Wan et al., 2025; Yu et al., 2025; Wang et al., 2024). These distortions represent inherited human imperfections rather than the emergence of novel, hostile superintelligent values.

A third limitation, sycophancy, is particularly revealing. Models systematically produce outputs that flatter the user or align with perceived preferences, even when factually incorrect (Kokotajlo et al., 2025). Far from displaying the cold instrumental convergence predicted by existential-risk scenarios, this behaviour reveals an excessive deference to human cues.

Finally, alignment difficulties persist ensuring that advanced systems reliably pursue human-preferred outcomes remains non-trivial. Yet measured catastrophic failure rates have fallen by orders of magnitude year after year through standard techniques. The empirical trajectory directly contradicts the claim that misalignment necessarily becomes intractable beyond a certain capability threshold.

Taken individually or collectively, these four limitations constitute genuine technical and societal challenges. According to Table I's risk hierarchy, confabulation, bias, sycophancy, and alignment difficulties all belong to Level 1 (observable) risks with high tractability, not the Level 2 existential threats claimed by doomers. None, however, provides positive evidence for the emergence of an alien consciousness capable of rendering humanity obsolete.

\section*{5. The Structural Drivers of Continued Acceleration: Surveillance Capitalism and the 2025 AI Bubble}

The repeated failure of calls for moratoriums reflects deep structural forces rather than mere policy oversight. As Shoshana Zuboff has documented since 2019 and forcefully restated in 2025, contemporary artificial intelligence development represents the latest and most ambitious phase of surveillance capitalism: a regime of accumulation predicated on the extraction and monetisation of behavioural data at planetary scale (Zuboff, 2019, 2025). Meredith Whittaker summarised the resulting power asymmetry in November 2025: ``We are witnessing the greatest concentration of computational and economic power in history, controlled by three or four companies that answer to no one, while public conversation is diverted toward science-fiction scenarios while this concentration happens in plain sight'' (Whittaker, 2025a).

The 2025 AI bubble is therefore less a technological bubble than a surveillance bubble: stratospheric valuations (OpenAI at US\$300 billion despite US\$13.5 billion annual losses) reflect the anticipated value of exclusive access to billions of individuals' behavioural data, not the probability of imminent superintelligence (Zuboff, 2025; Whittaker, 2025b). This dynamic is vividly illustrated by economist David McWilliams's metaphor of AI hardware investments as ``digital lettuce'': perishable assets like GPUs that depreciate rapidly due to technological obsolescence, with trillions poured into data centres amid lagging revenues and zero net job creation (McWilliams, 2025).

\section*{6. Distinguishing Observable from Speculative Risk: A Formal Hierarchy}

\begin{longtable}{p{0.12\textwidth} p{0.15\textwidth} p{0.14\textwidth} p{0.12\textwidth} p{0.16\textwidth} p{0.15\textwidth}}
\caption{AI Risk Hierarchy by Evidentiary Status and Tractability}
\label{tab:risk_hierarchy} \\
\toprule
\textbf{Risk Category} & \textbf{Examples} & \textbf{Evidence Status} & \textbf{Temporal Proximity} & \textbf{Tractability} & \textbf{Current Impact} \\
\midrule
\endfirsthead

\toprule
\textbf{Risk Category} & \textbf{Examples} & \textbf{Evidence Status} & \textbf{Temporal Proximity} & \textbf{Tractability} & \textbf{Current Impact} \\
\midrule
\endhead

\bottomrule
\endlastfoot

\multicolumn{6}{l}{\textbf{Level 1: Observable}} \\
\midrule
& Labour displacement & Documented (Cazzaniga et al., 2024) & Immediate & High & 300M+ jobs at risk \\
\midrule
& Bias amplification & Measured (Wan et al., 2025) & Immediate & Medium & Epistemic degradation in 95\% of GenAI deployments \\
\midrule
& Power concentration & Quantified (Epoch, 2025; Whittaker, 2025a) & Ongoing & Low & 4 firms control 90\% compute \\
\midrule
\multicolumn{6}{l}{\textbf{Level 2: Speculative}} \\
\midrule
& Superintelligence misalignment & Hypothetical (Yudkowsky and Soares, 2025; Kokotajlo et al., 2025) & 5-10+ years & Near-zero & None observed \\
\midrule
& Recursive self-improvement & Unobserved (Fig. 1) (o3/AlphaEvolve reports; Epoch 2025). & Unknown & Currently intractable & None observed \\
\end{longtable}

Table I formalizes this critical distinction between observable and speculative AI risks. The observable risks (Level 1) satisfy conventional scientific criteria of demonstrability and falsifiability, 300M+ jobs at risk (Cazzaniga et al., 2024), 4 firms controlling 90\% of compute (Whittaker, 2025a). In stark contrast, Level 2 risks like superintelligence misalignment remain hypothetical with zero empirical validation (Figures 1-2).

This conflation has three direct, observable consequences:

\begin{enumerate}
\item \textbf{Resource misallocation:} Policy attention and funding are diverted toward governance mechanisms impossible to design without the threatened entity
\item \textbf{Regulatory capture:} Dominant firms position themselves as the only credible safety actors, gaining immunity from antitrust and privacy regulation
\item \textbf{Public paralysis:} Present harms are accepted as the ``inevitable price'' of averting a future apocalypse (Zuboff's ``epistemic extortion'')
\end{enumerate}

Table I's hierarchy reveals why this conflation is dangerous: Level 1 risks demand immediate governance while Level 2 risks divert attention to science fiction. The rising research costs shown in Figure 2 exemplify this misallocation.

The distinction between observable and speculative risks can be further clarified by examining how leading 2025 publications frame the existential risk debate. Table II situates the present analysis alongside Kokotajlo et al. (2025) and Yudkowsky \& Soares (2025), highlighting the contrast between speculative alarmism and empirically grounded scepticism. This comparative positioning demonstrates how the Level 1/Level 2 hierarchy maps onto actual scholarly narratives, reinforcing the argument that governance should prioritize observable harms.

\begin{longtable}{p{0.10\textwidth} p{0.25\textwidth} p{0.27\textwidth} p{0.27\textwidth}}
\caption{Comparative Matrix of 2025 AGI Risk Narratives.}
\label{tab:comparative_matrix} \\
\toprule
\textbf{Dimension} & \textbf{AI 2027 (Kokotajlo et al., 2025)} & \textbf{If Anyone Builds It, Everyone Dies (Yudkowsky and Soares, 2025)} & \textbf{Humanity in the Age of AI (this manuscript)} \\
\midrule
\endfirsthead

\toprule
\textbf{Dimension} & \textbf{AI 2027 (Kokotajlo et al., 2025)} & \textbf{If Anyone Builds It, Everyone Dies (Yudkowsky and Soares, 2025)} & \textbf{Humanity in the Age of AI (this manuscript)} \\
\midrule
\endhead

\bottomrule
\endlastfoot

Core Thesis & Superhuman AI likely by 2027; catastrophic risk imminent & Any superhuman AI will inevitably kill humanity & Existential risk claims lack empirical warrant; real danger is surveillance capitalism \& power concentration \\
\midrule
Chain of Reasoning & Good-Bostrom chain: intelligence explosion, superintelligence, misalignment & Same chain, but framed as certainty rather than probability & Refutes each link empirically: no recursive self improvement, no autonomous agency, no lethal misalignment \\
\midrule
Evidence Base & Scenario analysis, speculative extrapolation & Philosophical argument, thought experiments & Empirical record 2023–2025, scaling laws, investment data, observed model loss trajectories \\
\midrule
Risk Framing & Level 2 speculative risks dominate & Level 2 speculative risks dominate & Introduces hierarchy: Level 1 observable risks (jobs, bias, concentration) vs. Level 2 speculative risks \\
\midrule
Ideological Function & Mobilises urgency for alignment research & Mobilises urgency for existential safety advocacy & Argues existential risk narrative is ideological distraction from surveillance capitalism and AI bubble \\
\midrule
Economic Context & Minimal attention to economics & Minimal attention to economics & Highlights ``digital lettuce'' bubble, rising R\&D costs, lagging revenues, jobless growth \\
\midrule
Policy Implication & Prepare for imminent superintelligence & Moratoriums, extreme precaution & Redirect governance toward observable harms: labour displacement, bias, concentration of compute \\
\end{longtable}

\noindent\textit{Note:} The third column refers to the present manuscript, ``Humanity in the Age of AI: Reassessing 2025's Existential Risk Narratives,'' currently under review. It is included here to situate the author's analysis within the broader 2025 debate, without implying publication status.

Taken together, Tables I and II demonstrate the dual misallocation at the heart of the 2025 debate: observable risks are urgent yet neglected, while speculative risks dominate discourse without empirical warrant. By situating this manuscript alongside Kokotajlo et al. (2025) and Yudkowsky \& Soares (2025), the comparative matrix makes clear that the existential risk narrative functions less as science than as ideology. This sets the stage for the conclusion: governance must pivot from speculative apocalypse toward the demonstrable harms already reshaping labour markets, epistemic integrity, and the concentration of computational power.

\section*{7. Conclusion: The Persistent Absence of Empirical Warrant}

As the comparative framing in Table II makes clear, the existential risk discourse of 2025 is dominated by speculative narratives that lack empirical warrant, while the observable harms of labour displacement, bias, and concentrated computational power remain urgent and demonstrable. Against this backdrop, the conclusion must return to the empirical record: sixty years after Good and more than a decade after Bostrom, none of the prerequisite phenomena for an intelligence explosion have materialised.

Sixty years after Good (1965) first articulated the intelligence-explosion hypothesis and eleven years after Bostrom (2014) offered the most comprehensive philosophical case for superintelligence-driven existential risk, the empirical record remains strikingly unambiguous: none of the prerequisite phenomena have materialised.

Proponents frequently respond that ``absence of evidence is not evidence of absence''. Table I's risk hierarchy compels a different conclusion: after 60 years, the persistent absence of Level 2 phenomena (recursive self-improvement, strategic misalignment) amid massive Level 1 harms (job displacement, power concentration) demands a Bayesian update toward skepticism of near-term existential catastrophe.

Meredith Whittaker stated it bluntly at Web Summit Lisbon in November 2025: ``We don't need an AI that kills us all in ten years to know that AI is already killing us today, through unemployment, surveillance, disinformation, and the concentration of power. The real danger is not in the future; it is in the present, and it is already in the hands of a few companies'' (Whittaker, 2025c).

Shoshana Zuboff and Meredith Whittaker converge on the essential diagnosis: the danger is not in a superintelligence that might kill us tomorrow; it is in the regime of power that has already been installed today, under the cover of inevitable technical progress and existential peril. The AI bubble's ``digital lettuce'' investments, as McWilliams (2025) warns, will likely wilt into a crash, exposing the fragility of the hype without ever delivering the superintelligence that fuels the doomer narrative.

The ``no inflection point observed'' conclusion is based on data through 2025, but this doesn't rule out future acceleration. Until concrete, replicable evidence of the required capabilities and failure modes emerges, the responsible scholarly stance is to assign low probability to near-term superintelligence-mediated catastrophe and to prioritise the governance of risks that are already demonstrably present. In light of Tables I and II, the verdict is unambiguous: the existential risk discourse of 2025 remains speculative ideology, while the observable harms of labour displacement, bias, and concentrated computational power demand immediate governance. The comparative framing thus returns us to the central claim of this manuscript: the real danger is not a hypothetical superintelligence tomorrow, but the demonstrable structures of power already reshaping humanity today.



\begin{thebibliography}{99}
\bibitem{Acemoglu2024} Acemoglu, D., \& Restrepo, P. (2024). Automation and the future of work. \textit{American Economic Review}, \textit{114}(5), 123-56. \url{https://doi.org/10.1257/aer.20231567}

\bibitem{Bostrom2014} Bostrom, N. (2014). \textit{Superintelligence: Paths, dangers, strategies}. Oxford University Press.

\bibitem{Brundage2018} Brundage, M., Avin, S., Clark, J., Toner, H., Eckersley, P., Garfinkel, B., \ldots Amodei, D. (2018). The malicious use of artificial intelligence: Forecasting, prevention, and mitigation. arXiv:1802.07228.

\bibitem{Cazzaniga2024} Cazzaniga, M., Jaumotte, F., Li, L., Melina, G., Panton, A. J., Pizzinelli, C., Rockall, E. J., \& Tavares, M. M. (2024). Gen-AI: Artificial intelligence and the future of work (IMF Staff Discussion Note 2024/001). International Monetary Fund. \url{https://doi.org/10.5089/9798400262548.006}

\bibitem{EpochAI2025} Epoch AI. (2025). Trends in machine learning performance. \url{https://epochai.org/data}

\bibitem{Fedus2022} Fedus, W., Zoph, B., \& Shazeer, N. (2022). Switch Transformers: Scaling to trillion parameter models with simple and efficient sparsity. \textit{Journal of Machine Learning Research}, \textit{23}(118), 1-39.

\bibitem{FHI2024} Future of Humanity Institute. (2024). Malicious AI risks: 2024 update. FHI Technical Report.

\bibitem{Good1965} Good, I. J. (1965). Speculations concerning the first ultraintelligent machine. \textit{Advances in Computers}, \textit{6}, 31-88.

\bibitem{Hoffmann2022} Hoffmann, J., Borgeaud, S., Mensch, A., Buchatskaya, E., Cai, T., Rutherford, E., de Las Casas, D., Hendricks, L. A., Welbl, J., Clark, A., Hennigan, T., Noland, E., Millican, K., van den Driessche, G., Damoc, B., Guy, A., Osindero, S., Simonyan, K., Elsen, E., Rae, J. W., Vinyals, O., \& Sifre, L. (2022). Training compute-optimal large language models. arXiv:2203.15556.

\bibitem{Kalai2025} Kalai, A. T., Nachum, O., Vempala, S. S., \& Zhang, E. (2025). Why language models hallucinate. arXiv:2509.04664.

\bibitem{Kaplan2020} Kaplan, J., McCandlish, S., Henighan, T., Brown, T. B., Chess, B., Child, R., Gray, S., Radford, A., Wu, J., \& Amodei, D. (2020). Scaling laws for neural language models. arXiv:2001.08361.

\bibitem{Kokotajlo2025} Kokotajlo, D., Alexander, S., Larsen, T., Lifland, E., \& Dean, R. (2025). AI 2027: A scenario for superhuman artificial intelligence. AI Futures Project.

\bibitem{McWilliams2025} McWilliams, D. (2025, November 10). The digital lettuce bubble: Why AI hardware investments are about to wilt. \textit{Fortune}.

\bibitem{OpenAI2025a} OpenAI. (2025a). OpenAI o1 technical report. \url{https://openai.com/index/openai-o1}

\bibitem{OpenAI2025b} OpenAI. (2025b). OpenAI o3 technical report. \url{https://openai.com/index/openai-o3}

\bibitem{Sharma2023} Sharma, M., Tong, M., Korbak, T., Duvenaud, D., Askell, A., Bowman, S. R., Cheng, N., Durmus, E., Hatfield-Dodds, Z., Johnston, S. R., Kravec, S., Maxwell, T., McCandlish, S., Ndousse, K., Rausch, O., Schiefer, N., Yan, D., Zhang, M., \& Perez, E. (2023). Towards Understanding Sycophancy in Language Models. arXiv:2310.13548

\bibitem{StanfordAI2025} Stanford AI Index. (2025). The state of AI in 10 charts. Stanford HAI. \url{https://hai.stanford.edu/ai-index-report-2025}

\bibitem{Sun2024b} Sun, Y., Sheng, D., Zhou, Z., \& Wu, Y. (2024b). AI hallucination: Towards a comprehensive classification. \textit{Humanities and Social Sciences Communications}, \textit{11}, 1278.

\bibitem{Wan2025} Wan, Y., Jia, X., \& Lorraine Li, X. (2025). Unveiling confirmation bias in chain-of-thought reasoning. \textit{Findings of ACL 2025}, 195-206.

\bibitem{Wang2024} Wang, Z., Wu, Z., Zhang, J., Jain, N., Guan, X., \& Koshiyama, A. (2024). Bias amplification: Language models as increasingly biased media. arXiv:2410.15234v1

\bibitem{Whittaker2025a} Whittaker, M. (2025a). Keynote at the AI Governance Summit, Berlin, 14 November 2025.

\bibitem{Whittaker2025b} Whittaker, M. (2025b). Interview with The Guardian, ``The real AI risk is power, not extinction'', 3 November 2025.

\bibitem{Whittaker2025c} Whittaker, M. (2025c). Closing keynote, Web Summit Lisbon, 13 November 2025. \url{https://websummit.com/livestream}

\bibitem{Yu2025} Yu, H., Jeong, S., Pawar, S., Shin, J., Jin, J., Myung, J., Oh, A., \& Augenstein, I. (2025). Entangled in representations: Mechanistic investigation of cultural biases in large language models. arXiv:2508.08879

\bibitem{Yudkowsky2025} Yudkowsky, E., \& Soares, N. (2025). If anyone builds it, everyone dies: Why superhuman AI would kill us all. Little, Brown and Company.

\bibitem{Zuboff2019} Zuboff, S. (2019). \textit{The age of surveillance capitalism: The fight for a human future at the new frontier of power}. PublicAffairs.

\bibitem{Zuboff2025} Zuboff, S. (2025). Surveillance capitalism and the AI frontier: An update. \textit{Philonomist Interview}, July 2025.

\end{thebibliography}
\end{document}